\documentclass{emulateapj}
\usepackage{amsmath}
\usepackage{natbib}
\usepackage{color}
\usepackage{soul}

\newcommand{\msun}{{M}_{\odot}}

\newcommand{\Lacc}{L_{\mathrm{acc}}}

\newcommand{\nudot}{\dot{\nu}}
\newcommand{\Jdot}{\dot{J}}
\newcommand{\JdotJ}{\frac{\dot{J}}{J}}
\newcommand{\Mdot}{\dot{M}}
\newcommand{\Porb}{P_{\mathrm{orb}}}
\newcommand{\Porbi}{P_{\mathrm{orb,i}}}
\newcommand{\mti}{M_{2,\mathrm{i}}}
\newcommand{\rti}{R_{2,\mathrm{i}}}
\newcommand{\moi}{M_{1,\mathrm{i}}}
\newcommand{\RL}{R_L}

\newcommand{\taumd}{\tau_{\dot{M}}}

\newcommand{\ee}[2]{\ensuremath{#1 \times 10^{#2}}}

\shorttitle{AM CVn Mass Transfer Turn-On: Implications for RX J0806 \& RX J1914 }
\shortauthors{Deloye \& Taam}

\begin{document}

\title{The Turn-On of Mass Transfer in AM CVn Binaries: Implications for RX J0806+1527 and RX J1914+2456} 
\author{Christopher J. Deloye \& Ronald E. Taam}
\affil{Department of Physics \& Astronomy, Northwestern University, 2131 Tech Drive, Evanston, IL 60208 }
\email{cjdeloye@northwestern.edu, r-taam@northwestern.edu}

\begin{abstract}
We report on evolutionary calculations of the onset of mass transfer in AM CVn binaries, treating the donor's evolution in detail.  We show that during the early contact phase, while the mass transfer rate, $\Mdot$, is increasing, gravity wave (GW) emission continues to drive the binary to shorter orbital period, $\Porb$.  We argue that the phase where $\Mdot > 0$ and $\nudot > 0$ ($\nu = 1/\Porb$) can last between $10^3$ and $10^6$ yrs, significantly longer than previously estimated.  These results are applied to RX J0806+1527 ($\Porb = 321$ s) and RX J914+2456 ($\Porb=569$ s), both of which have measured $\nudot > 0$.  \emph{Thus, a $\nudot > 0$  does not select between the unipolar inductor and accretion driven models proposed as the source of X-rays in these systems}.  For the accretion model, we predict for RX J0806 that $\ddot{\nu} \approx \ee{1.0-1.5}{-28}$ Hz s$^{-2}$ and argue that timing observations can probe $\ddot{\nu}$ at this level with a total $\approx 20$ yr baseline. We also  place constraints on each system's initial parameters given current observational data.
\end{abstract}


\section{Introduction \label{section:intro}}
The true nature of the two candidate ultracompact binary systems originally discovered with \textit{ROSAT},  RX J0806+1527 \citep{beuermann99} and RX J1914+2456 \citep{haberl95}, has been a source of much controversy. Both systems are soft X-ray sources whose light curves are modulated on periods of 321 s (RX J0806, hereafter J0806) and 569 s (RX J1914, hereafter J1914) \citep{motch96,israel99,burwitz01}. In both cases, the X-ray light is 100\% modulated for roughly half the period \citep{cropper98,israel99} and the optical light is modulated on the same period as the X-ray light \citep{ramsay00,israel02,ramsay02a,ramsay02b,israel03} with little evidence for other periodicities \citep[but see][]{ramsay06}. It is widely believed, based partially on the stability and singularity of these periods, that the modulations are on the systems' orbital periods, $\Porb$ \citep[although see][for an alternate interpretation]{norton04}. This would make J0806 and J1914 the two shortest period binaries known.


The X-ray production mechanism in these systems has been much debated. Two competing models have come to the fore: the unipolar inductor (UI) model  \citep{wu02} and the direct-impact accretion model \citep{marsh02}. Both model these systems as white dwarf-white dwarf (WD-WD) binaries with $\Porb$ equal to the X-ray period. In the UI model, a magnetized, more massive primary WD spinning asynchronously with respect to the orbit induces an electric field in the secondary WD and drives a current between the two. X-rays are produced by resistive dissipation in the primary's atmosphere.  In the accretion model, the secondary fills it Roche Lobe ($\RL$) and mass transfer occurs. The binary's compact geometry leads to the accretion stream directly impacting the primary and no accretion disk forms. Both models lead to a spatially small X-ray production site on the primary, explaining, in principle, the observed  X-ray modulation \citep[see, however][]{barros05}.

One potential means of distinguishing between the two models is the \emph{secular} time derivative of $\Porb$, or equivalently of the orbital frequency, $\nu = 1/\Porb$, $\nudot$. The UI model predicts $\nudot > 0$. Assuming fully degenerate donors and enforcing exactly that the donor's radius, $R_2$ equals its Roche radius: $\RL = R_2$, the accretion model would predict $\nudot < 0$.  Determinations of $\nudot$ from both optical \citep{hakala03,hakala04,ramsay05} and X-ray observations \citep{strohmayer02,strohmayer03,strohmayer04,strohmayer05} \emph{all} show that $\nudot > 0$ over the entire epoch of observations. On the assumption that these results reflect the \emph{secular} $\nudot$, this has been taken as evidence against accretion models \citep[e.g.;][]{israel03,strohmayer02,strohmayer04,ramsay05}.  However, the $R_2 = \RL$ constraint requires a very large mass transfer rate $\Mdot \approx 10^{-7} - 10^{-5} \msun$ yr$^{-1}$ and if $\Mdot$ is significantly below this ``equilibrium'' value, observing a $\nudot > 0$ is possible.  The $\Mdot$ could be lower than expected either due to a non-secular mechanism \citep{marsh05} or during the mass transfer turn-on phase \citep{willems05}.  However, the timescales on which the binary (re)establishes its equilibrium $\Mdot$ are estimated to be very short: 100-1000 yrs \citep{marsh05} or 2-20 yrs \citep{willems05}, making observing such phases unlikely. 

Here we reexamine the suggestion J0806 and J1914 are direct-impact accretors seen during the $\Mdot$ turn-on phase in the context of stellar evolution models developed to realistically address the complete evolution of AM CVn binaries.  In \S 2, we show that the population of AM CVns forming through a double-degenerate channel \citep[see][]{nelemans01,deloye05} naturally produces systems whose contact phase evolution give $\Porb$ and $\nudot$ consistent with both J0806 and J1914. The maximum $\Mdot$ is sufficiently high in these systems that $\RL$ must penetrate below the donor's photosphere, naturally increasing the $\Mdot$ growth times to $\approx 10^2-10^6$ yrs.  We explore in \S 3 the range of initial conditions that will lead to J0806 and J1914. We determine constraints on these initial conditions and present predictions for $\ddot{\nu}$ in J0806 as a diagnostic tool.  Finally, we present our discussion in \S 4.
 
\section{Mass Transfer Turn-on in AM CVn Systems\label{section:parameter_space}}
We take as our model for J0806 and J1914 a WD-WD binary with primary and secondary (donor) masses $M_1$ and $M_2$, where the donor is pure He. Binary evolution through two common envelope (CE) phases can produce such systems that gravity wave (GW) emission can drive into contact within a Hubble time \citep{nelemans01}. During this pre-contact phase, the donor cools and contracts.  Variations in the donor's pre-CE entropy and time to contact lead to a range in the donor's degeneracy and radius, $R_2$, at contact \citep{deloye03,deloye05}. For fixed $M_2$, less degenerate donors have larger $R_2$, leading to longer $\Porb$ at contact, $\Porbi$.

We parameterize the system's state at contact by $\moi$, $\mti$, and $\rti$.  The range of $\moi$, $\mti$ pairs we consider are taken from \citet{nelemans01}.  The $\rti$ range is determined using the prescription outlined in \citet{deloye05} with slight modifications detailed in a companion paper (Deloye et al. 2006, \textit{in preparation}).  For this population, fully degenerate donors produce $\Porbi \leq 6$ min; a realistic treatment of pre-contact donor cooling gives $\Porbi$ extending well above 10 min \citep{deloye05}.  \emph{Thus an accretion model can easily accommodate the $\Porb$ of both J0806 and J1914 as early contact systems}.

We have carried out a range of evolutionary calculations beginning from the initial conditions determined above.  We utilized a coupled stellar/binary evolution code specifically developed for these calculations and followed each model's evolution from pre-contact through the late stages of the mass transfer phase. We assume conservative mass transfer, circular orbits, and that GW emission is the only source of orbital angular momentum, $J$, loss. We use the prescription of  \citet{ritter88} to calculate $\Mdot$ when $\RL \geq R_2$ and adopt the prescription of \citet{kolb90} when $\RL < R_2$.  

\begin{figure}
\plotone{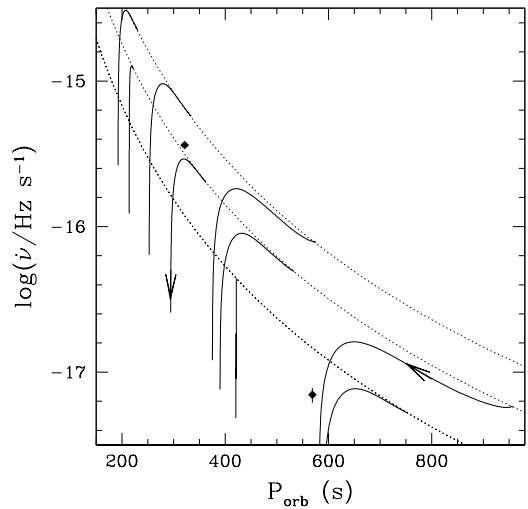}
\caption{Comparison between the predicted, secular evolution of $\nudot$ vs. $\Porb$ and the measured values of $\nudot$ in RX J0806 and RX J1914 \citep{strohmayer04,strohmayer05}.  Solid lines show our full evolutionary calculations for a set of initial conditions representative of those expected in the pre-AM CVn population.  Arrows on several of the tracks indicate the sense of evolution. The dotted lines show the contribution of the GW term (i.e. $\Mdot=0$) in equation (\ref{eq:nudot}) for the same $\moi$, $\mti$ as the full calculations. The three lines have, from bottom to top in $\msun$: ($\moi$, $\mti$) = (0.4,0.1), (0.625,0.2), and (1.025,0.3).  The different solid lines starting on each dotted line differ in initial entropy (and hence $\rti$).  The diamonds with error bars (only visible in J1914) show the observed locations of $\nudot$ for J0806 and J1914. \label{fig:nudot_Porb}}
\end{figure}

Under these assumptions, $\nudot$ is given by
\begin{equation}
\nudot = -3 \nu \left[ \JdotJ - (1-q) \frac{\Mdot_2}{M_2} \right]\,, \label{eq:nudot}
\end{equation}
where $\Jdot/J$ is given in \citet{landau71}, $q = M_2/M_1$, and $\Mdot_2 = -\Mdot < 0$.  As $\Jdot/J < 0$, the first term in equation (\ref{eq:nudot}) drives $\nudot > 0$; a sufficiently high $\Mdot$ is required for $\nudot \leq 0$. After the donor comes into contact, there is a period of time in which both $\Mdot_2 \neq 0$ and $\nudot > 0$. How quickly $\nudot$ reaches 0 depends on the growth rate of $\Mdot$. 

To place J0806 and J1914 in the context of the turn-on phase in AM CVn binaries, we show in Figure \ref{fig:nudot_Porb} the $\nudot$-$\Porb$ evolution for a sample of our calculations.  Solid lines originating on the same dotted line share the same $\moi$, $\mti$, but different $\rti$.  The dotted lines show the corresponding relation for GW losses alone. The diamonds show the measured $\nudot$ for J0806 \citep[$\ee{3.63}{-16}$ Hz s$^{-1}$;][]{strohmayer05} and J1914  \citep[$\ee{7}{-18}$ Hz s$^{-1}$;][]{strohmayer04}. \emph{The set of initial conditions expected for double-degenerate channel AM CVn progenitors naturally produce secular evolution consistent with the observations of both systems}. While J0806 is consistent with a system whose $\nudot$ is dominated by GW losses, J1914 is only so for very low values of $\moi$, $\mti$: at $\Mdot = 0$, if $\moi> 0.15$, $\mti <0.1$ is required. The combination of these low masses are not plausible outcomes of binary evolution \citep{nelemans01}. Thus, regardless of X-ray production model, a significant negative contribution to  $\nudot$ (e.g., due to mass transfer or properly oriented spin-orbit coupling, as in the UI model) is \emph{required} for J1914.  In an accretion model, this means the $\Mdot$ in J1914 must be large.

While for $\RL > R_2$, $\Mdot$ grows exponentially with $\Delta R = \RL-R_2$ \citep{ritter88}, this growth rate slows considerably once $\RL < R_2$ \citep{kolb90}.  The donors also initially have non-degenerate outer layers that, in most cases, are predominantly radiative, producing contraction upon mass loss during the $\Mdot$ turn-on phase.  Sub-photospheric mass transfer and donor contraction both increase the duration of the turn-on phase beyond the previous estimate of $\taumd = \Mdot/\ddot{M} \sim 1-10$ yrs made for J0806 at $\Mdot=0$ \citep{willems05}.  In our full calculations,  systems with $\Porb < 10$ min reach $\nudot =0$ after $10^3-10^6$ yrs. Thus, \emph{AM CVn binaries remain in the $\Mdot$ turn-on phase significantly longer than previously estimated.}  

\section{Accretion Model Implications \label{section:systems}}
We have shown that a secular $\nudot>0$ does not preclude accretion driven models for J0806 and J1914.  We now discuss the implications of accretion models for these systems.   We first determine $\moi$, $\mti$, and $\rti$ leading to systems consistent with the observed $\Porb$ and $\nudot$ in J08086 and J1914 and use the successful initial conditions to determine expected ranges for current system properties.

\begin{figure}
\plotone{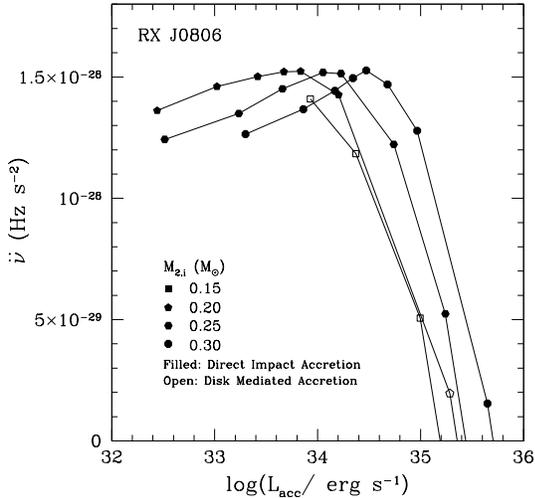}
\caption{The predicted $\ddot{\nu}$ versus $\Lacc$ from our full evolutionary calculations for systems with the measured $\nudot$ and $\Porb$ of RX J0806.  Symbol shapes indicate the initial donor mass, $M_{2,i}$ as indicated in the Figure's key. The solid lines are included to help discern lines of constant $M_{2,i}$.  Filled symbols indicated systems that have always accreted via direct-impact.
\label{fig:j0806_nddot_lacc}}
\end{figure}

We first consider $\ddot{\nu}$ in J0806.  For GW emission alone, this is given by 
\begin{equation}
\ddot{\nu}_{\mathrm{GR}} = \ee{4.95}{-30}  \mathrm{Hz\, s^{-2}}\, \left(\frac{\nu}{10^{-3}\, \mathrm{Hz}}\right)^{19/3} \left(\frac{\mathcal{M}}{0.871 \msun}\right)^{10/3}\,, \label{eq:nuddot}
\end{equation}
where $\mathcal{M}^{5/3} = (M_1 M_2)/(M_1+M_2)^{1/3}$ is the system's chirp mass. For $\Mdot > 0$,  $\ddot{\nu} < \ddot{\nu}_{\mathrm{GR}}$ generically. For models passing through  $\Porb = 321.5$ and $\nudot = \ee{3.63}{-16}$ Hz s$^{-1}$, we calculate $\ddot{\nu}$ and $\Lacc = \Mdot (\phi_{\mathrm{L1}} - \phi_{R_1})$, where $\phi_{\mathrm{L1}}$, $\phi_{R_1}$ are the potential at the inner Lagrange point and surface of the accretor \citep{han99}.  We plot these predictions in Figure \ref{fig:j0806_nddot_lacc}.  The lines connect models with equal $\mti$ and larger $\Lacc$ values correspond to larger $\rti$. For low $\Mdot$, larger $\rti$ requires larger $\mathcal{M}$ to satisfy the $\nudot$ constraint.  As $\rti$ increases, $\Mdot$ contributions to $\ddot{\nu}$ become important, reducing $\ddot{\nu}$ and eventually driving it negative.

For J0806, the observed X-ray luminosity is $L_X \approx \ee{1-5}{32}$ erg s$^{-1}$ \citep[for a distance of 500 pc;][]{israel99,israel03}. Given uncertainties in the source distance and the conversion efficienct of accretion energy into soft X-rays, we take as a rough upper limit for $\Lacc \approx 10^{34}$ erg s$^{-1}$. This places J0806 in the regime where $\Mdot$ contributions to $\ddot{\nu}$ are of order or less than GW contributions. We predict $\ddot{\nu} = \ee{1-1.5}{-28}$ Hz s$^{-2}$. Given a phase measurement accuracy of $\sim 0.01$, a total timing measurement baseline of 20 yrs should be sufficient to constrain $\ddot{\nu}$ at this level (Strohmayer 2006, \textit{private communication}).  Thus, this prediction should be testable within 5-10 yrs.

We indicate by solid symbols in Figure \ref{fig:j0806_nddot_lacc} systems in which accretion is via direct impact.  Systems which form a disk (open symbols) are unlikely models for J0806 since it is then unclear how the 100\% modulation of the X-ray light would be obtained.  The direct-impact constraint requires $\mti \gtrsim 0.20 \msun$ in J0806.  We also check if the advection of orbital angular momentum onto the accretor is significant \citep{marsh04}.  We compare the GW $\Jdot/J$ to $\sqrt{(1+q) r_h} \Mdot_2/M_2$, where $r_h$ is the effective radius at which an orbit has the same specific angular momentum as the transferred matter \citep[see][]{verbunt88,marsh04} and find that for $\Lacc < 10^{34}$ erg s$^{-1}$ advection contributies $<1\%$ to the $J$ evolution.  Finally for the direct-impact models, $\taumd \approx 10^2-10^4$ yrs (increasing with $\Lacc$). 

For J1914, the small $\nudot$ requires a large $\Mdot \gtrsim \ee{3}{-9} \msun$ yr$^{-1}$.  At its longer $\Porb$, the phase space leading to direct-impact systems is small: we find that $\mti \approx 0.15-0.20 \msun$ is required and that $\rti$ must be close to the smallest value producing systems consistent with J1914.  This $\mti$ range is that most likely produced by binary evolution \citep{nelemans01}, so a fine-tuning problem is somewhat alleviated. The potential contribution of $J$ advection onto the accretor is significant in J1914: $( \sqrt{(1+q) r_h} \Mdot_2/M_2\; J/\Jdot ) \approx 0.4-0.6$.  Given the large uncertainty in tidal synchronization rates that would feed this $J$ back into the orbit \citep[see][and references therein]{marsh04}, we leave this point for future investigation.  Our direct-impact accreting models have $\Lacc \approx 10^{34}-10^{35}$ and  $\taumd \approx \ee{5}{5}$ yrs, consistent with the estimates for $L_X \approx 10^{33}-10^{35}$ erg s$^{-1}$ \citep{ramsay05,ramsay06}.

\begin{figure}
\plotone{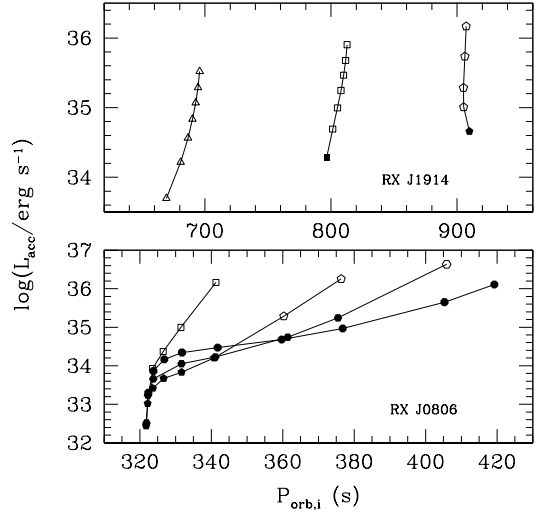}
\caption{The relation between $\Lacc$ at the point each model reaches the $\Porb$ of J0806 (lower panel) or J1914 (upper panel) and that model's initial $\Porb$ for models with $\nudot$ equal to the observed value is each system.  The symbols have the same meaning as in Figure \ref{fig:j0806_nddot_lacc} with the addition that triangles show models with $M_{2.i} = 0.1 \msun$.  The much narrower $\Porbi$ range in J1914 at fixed $\mti$ is due to the steep dependence of $\nudot$ on $\Porb$ once $\Mdot$ is large enough to compete with GW emission in setting $\nudot$ (see Figure \ref{fig:nudot_Porb}) \label{fig:Poi_Lacc}}
\end{figure}
With the assumption that any $J$ advection is negligible or is cancelled by sufficiently strong tidal coupling, we can place constraints on initial conditions producing systems consistent with J0806 and J1914. In Figure \ref{fig:Poi_Lacc}, we show our constraints on $\Porbi$ vs. $\Lacc$.

\section{Summary \& Discussion \label{section:summary}}
We have shown that the measured $\nudot$ in J0806 and J1914 are consistent with the \emph{secular} $\nudot$ expected for accretion driven systems leading to AM CVn binaries.  As such, the $\nudot$ measurements \emph{do not} provide a means of distinguishing between accretion and UI models posited for these systems.  We have shown that accretion models consistent with observations give $\taumd \approx 10^2-10^4$ yrs in J0806 and $\taumd \approx \ee{5}{5}$ yrs in J1914. For comparison, \citet{dallosso06} estimate that the UI timescale for J1914 is $\approx 10^4 - \ee{3}{5}$ yrs. We also predicted that accretion models give $\ddot{\nu} \approx \ee{1-1.5}{-28}$ Hz s$^{-2}$ for J0806; another 5-10 yrs baseline of timing J0806 will be sufficient to test this prediction. Quantitative predictions of $\ddot{\nu}$ for the UI model are needed to determine if $\ddot{\nu}$ will be a useful diagnostic tool. Our calculations also constrain each system's initial conditions and we relate the current $\Lacc$ to $\Porbi$ in both J0806 and J1914.

Recenlty, \citet{dantona06} proposed that J0806 could be modeled with a degenerate He WD donor with a thick H atmosphere supporting $p-p$ burning. In their model, $\Porb$ decreases until the H layer is removed, exposing the degenerate He WD.  Our pure He donor models shows that a thick H layer is not necessary to explain a phase where $\nudot > 0$ during early-contact AM CVn evolution. The combination of sub-photospheric mass transfer and spread in initial donor degeneracy are sufficient to produce either J0806 and J1914.  Indeed, J1914 is naturally accomodated within the AM CVn initial conditions of our model, but not by \citet{dantona06}, presumably becuase they assume a fully degenerate donor interior. We agree, however, that the presence of H in J0806 or J1914 is not unexpected given the uncertainty in the amount of H left on the donor after the CE phase.  Pure He models are a limiting case that could be altered to include an arbitrarily thick H layer, changing our results quantitatively. As such, a detection of H in J0806 or J1914 does not rule out accretion models nor on its own discriminate between our and the \citet{dantona06} models. 

Future work on these systems must seek to diagnose the source of their X-ray light.  An avenue for this would be consideration of the optical output of the systems where we would expect significant differences between the UI and accretion models. Assessment of potential fine-tuning problems for accretion models in J1914 is also needed.  This would be best performed in a general study addressing the detection probability for early contact AM CVns. Semi-degenerate systems are formed less frequently \citep{deloye05}, but evolve more slowly at longer $\Porb$, so the \textit{a prioi} expectations are unclear.  This would also constrain theoretical expectations for the intial conditions leading to AM CVn binaries. The usefulness of such studies, however, hinges on the determination that J0806 and J1914 are indeed direct-impact accretors, not unipolar inductors.  The community should continue efforts toward distinguishing between these models for these enigmatic systems.

\acknowledgements
We would like to thank Tod Strohymayer and Bart Willems for helpful discussions and feedback related to this work, as well as Gijs Roelofs and Susana Barros for their encouragement during its preparation. We also would like to thank Tom Marsh and Gijs Nelemans for providing comments on drafts of this paper.  This work was partially supported by the National Science Foundation through grant AST 02-00876.

\end{document}